\documentclass[reprint,amsmath,amssymb, aps,]{revtex4-2}

\usepackage{graphicx}
\usepackage{dcolumn}
\usepackage{bm}
\usepackage{amsmath}
\usepackage{amssymb}
\newcommand{\dg}{${^{\circ}}$}
\begin{document}

\preprint{APS/123-QED}

\title{Growth of rare-earth monopnictide DySb single crystal by novel Self-flux method}

\author{Mukesh Kumar Dasoundhi}
\author{Archana Lakhani}%
 \email{archnalakhani@gmail.com}
\affiliation{UGC-DAE Consortium for Scientific Research, University Campus, Khandwa Road, Indore-452 001, India}%

\date{\today}

\begin{abstract}
This report  presents a new synthesis protocol for the single crystal growth of rare earth monopnictide DySb by self-flux technique. A detailed structural, transport and magnetic characterisation have been done using X-Ray diffraction (XRD), High resolution X-Ray diffraction (HRXRD), resistivity and magnetisation measurements respectively. The Rietveld refinement of powder XRD pattern confirms that the grown crystal is in single phase and crystallizes in space group Fm$\bar{3}$m (225) of rock-salt type crystal structure. HRXRD on cleaved crystal confirms the single crystalline nature while rocking curve analysis reveals the high quality of the grown crystal. Temperature dependent resistivity and magnetisation measurements show a transition at 9.7\,K from paramagnetic (PM) to antiferromagnetic (AFM) state.
\end{abstract}

\keywords{Single crystal, Self flux, crystal growth, magnetic transition.}
\maketitle

\section{Introduction}
Recently, the rare-earth monopnictides (R-Sb/Bi) group have fascinated the researchers due to their novel electronic band structures. The novel electronic band structure in R-Sb/Bi leads to the observation of variety of exotic phenomena like non-saturating extremely large magnetoresistance (XMR), field-induced resistivity upturn with plateau, high carrier mobility, topological surface states and quantum oscillations\cite{1,2,3,4,5}.These salient features have raised immense interest in both fundamental understanding as well as in applied physics. XMR and field induced resistivity upturn with plateau are the bookmark phenomena of R-Sb/Bi which is quite similar to other trivial and non-trivial semimetals\cite{6,7,8,9}. The observation of ultrahigh mobility and XMR of $\sim$10$^{6}$\% in LaSb at T=2\,K and H= 9\,T, have triggered immense interest in these systems\cite{1}. Subsequently, similar results have been reported in other R-Sb/Bi and shown the non-trivial nature of band structure\cite{3,4,5,10}. DySb is a unique member among the R-Sb/Bi family due to its peculiar structural, magnetic and magnetotransport properties particularly at low temperatures. DySb hosts rock-salt type crystal structure having space group Fm$\bar{3}$m (225) at room temperature which transforms to a tetragonal structure with space group I4/mmm (139) around 10\,K\cite{11,12}. Accompanying the structural phase transition, a simultaneous magnetic transition occurs from paramagnetic (PM) to antiferromagnetic (AFM ) phase\cite{5,11,12,13,14}. The magnetic ground state can be altered dramatically with the application of magnetic field in different crystallographic directions. Recently Lie et al. have shown field-induced multiple phases and tricritical phenomena (boundary of antiferromagnetic, forced ferromagnetic and paramagnetic phase) in DySb single crystal\cite{12}. Hu et al. have reported a large magneto-entropy change of -20.6\,J/kgK for a field change of 7\,T across the magnetic transition (T$_{N}$=11\,K) in polycrystalline DySb while Li et al. have shown -36.4\,J/kgK of magneto-entropy change for a field change of 5\,T in single crystalline DySb\cite{14,15}.  Liang et al. have observed XMR in single crystalline DySb which is attributed to very high mobility of charge carriers arising from linear dispersion of bands \cite{5}. Therefore, the field-induced magnetic phase transitions with peculiar carrier dynamics make it an ideal candidate for magnetic topological semimetal. In order to exploit the maximum XMR in DySb and its topological character, it is all the more important to have minimal grain boundary effects on its carrier dynamics. There are various reports in literature on the growth of DySb crystals, most common being the flux method. For DySb, Sn flux method is used, where excess flux is removed at sufficiently high temperature\cite{5,12}. The DySb crystals used by Fetcher et al.  and Li et al. in their studies have been grown by Bridgeman method at high annealing temperature of $\sim$2500\dg\,C\cite{11,15} while Everett et al. have grown their crystals by fractional distillation of constituent matrix followed by vapor transport in a tungsten crucible\cite{13}. In order to obtain high quality single crystal for obtaining the best characteristic features of this material we adapted alternate techniques for the synthesis of DySb single crystals.\par
In this paper, we report our maiden effort on the successful growth  of DySb single crystal by ‘self-flux’ technique. The phase purity and single crystalline nature of grown crystal is systematically characterized by XRD and HRXRD respectively, while the phase transition behavior is probed by electrical transport and magnetisation measurements.

\section{Experimental}

\begin{figure}
	\centering
	\includegraphics[width=1.0\linewidth,height=0.6\linewidth]{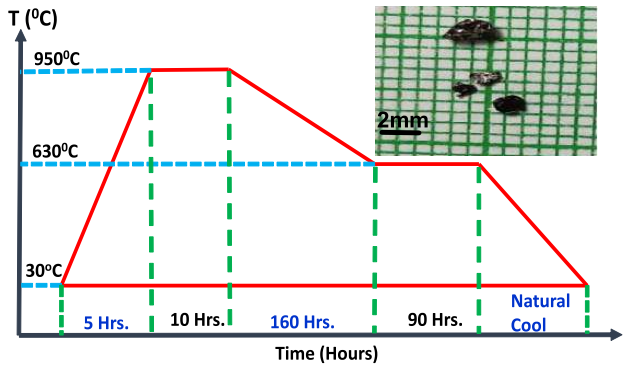}
	\caption{Time-Temperature profile for the growth of DySb single crystal, along with cleaved crystals.}	
\end{figure}

Single crystal of DySb has been prepared by self-flux method, which is quite different from earlier reported methods for the growth of DySb single crystal\cite{11,12,13,15}. Here, Dy (99.9\%) and Sb (99.999\%) granules are taken in 1:19 molar ratio, where the excess Sb acts as flux. As the constituent material (Sb) itself acts as a flux, it is termed as self-flux. The starting materials are homogeneously mixed for several hours using pestle and mortar. Thereafter, homogenously mixed powder of constituent elements are vacuum sealed ($\sim$10$^{-6}$ \,torr) in a quartz tube. The sealed quartz tube is placed in a programmable furnace, where it is heated upto 950\dg C and dwelled for 10 hours. The furnace is then cooled down to 630\dg C at the rate of 2\dg C/hrs., where it is dwelled for 90 hrs. followed by natural cooling. The sealed quartz tube is carefully removed from the furnace and then broken in order to remove the grown crystal. Figure 1 illustrates the schematic diagram of heat treatment protocol for DySb single crystal growth along with the photograph of grown crystals. After mechanically removing the Sb flux and cleaving, tiny shiny crystals of DySb ranging from 0.5 mm-2 mm are obtained. Some of the cleaved crystals are crushed and homogeneously powdered in the mortar with the help of pestle for powder X-Ray diffraction (XRD) studies. The powder XRD spectra were recorded at room temperature by Bruker D8 Advance diffractometer using  Cu-K$_\alpha$ source. The XRD data is recorded in the range of 10\dg-100\dg with step size of 0.02\dg. The obtained powder XRD data is fitted with FULLPROF software and then unit cell structure is generated using VESTA software. For determining the crystalline quality High Resolution X-Ray Diffraction (HRXRD) measurements ($\phi$-scan and $\omega$-scan) are performed in Bruker D8 Discover Diffractometer. Electrical transport and magnetisation measurements are performed in Physical Property Measurement System in the temperature range of 300\,K to 2\,K.

\section{Results and Discussions}

\subsection{Structural Characterization}

\begin{figure}
	\centering
	\includegraphics[width=1.05\linewidth,height=0.85\linewidth]{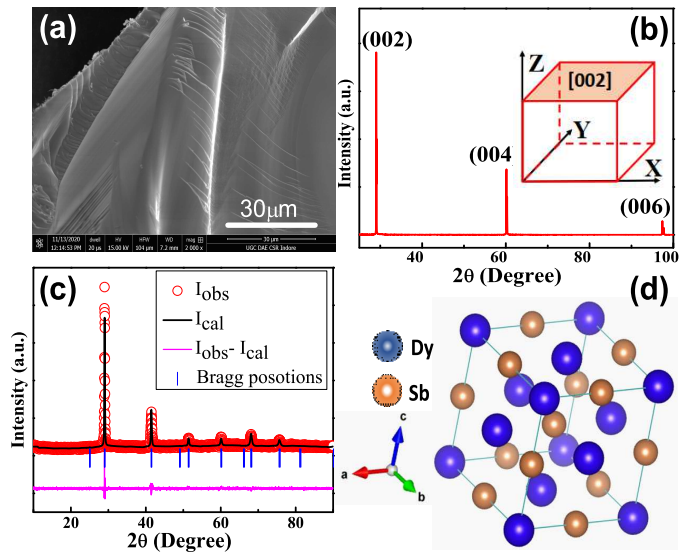}
	\caption{(a) FESEM image of cleaved DySb crystal. (b) $\theta$-2$\theta$, XRD scan of $\{$002$\}$ planes of cleaved surface. Inset shows the schematic of cubic crystal with [002] oriented. (c) Powder XRD pattern of DySb crystal with Rietveld refinement. Red circles represent the experimental data, black solid line is calculated intensity, blue bar shows Bragg peak position and solid pink line is the difference between experimental and calculated data. (d) Unit cell structure of DySb, generated from VESTA software.}
\end{figure}

\begin{table*}[ht]
	\caption{\textbf{Rietveld refined parameters}}
	\centering
	\begin{tabular}{p{0.10\linewidth}p{0.10\linewidth}p{0.10\linewidth}p{0.10\linewidth}p{0.10\linewidth}p{0.04\linewidth}}
		\hline
		$\alpha$=$\beta$=$\gamma$ & a=b=c & Volume & Wyckoff & Occupancy & $\chi$$^2$ \\ 
		& (\AA)& (\AA$^3$) & Position \\ 
		\hline 
		90$^o$ & 6.155(1) & 233.27(4) & 4a(0,0,0), & Dy-1.0 & 1.35\\ 
		 & & & 4b($\frac{1}{2}$,$\frac{1}{2}$,$\frac{1}{2}$) & Sb-0.9  \\
		\hline
	\end{tabular}
\end{table*}

Figure 2(a) displays the FESEM image of cleaved crystal. Figure 2(b) shows the out-of-plane XRD pattern of cleaved DySb crystal along (002) plane direction. The appearence of very sharp $\{$002$\}$ set of planes suggest that the grown crystal is oriented along [002] plane, which is further characterised by HRXRD measurements. Inset represents a schematic of cubic crystal showing [002] is the out-of-plane direction as observed in XRD. DySb crystallises in NaCl-type crystal structure in cubic space group Fm$\bar{3}$m (225)\cite{5,11,12}. The lattice parameter "a" calculated from Bragg's law (2d$\sin$$\theta$=n$\lambda$) for the observed out-of-plane XRD pattern is a=6.15(4)\AA. The calculated lattice parameter matches well with the JCPDS (15084) value of a=6.153\AA\ for DySb crystal as well as with the reported values of a=6.143\AA\cite{11} and a=6.161\AA \cite{14}. Figure 2(c) displays the Rietveld refinement fit to the powder XRD pattern of crushed DySb crystal. All the peaks are  well indexed and fitted on the basis of NaCl-type crystal structure with Fm$\bar{3}$m (225) space group. The absence of extra peaks in powder XRD pattern confirms the single-phase nature of grown crystal. The refined lattice parameters are a=b=c=6.155(1)\AA, which resembles to the calculated lattice parameters for the out-of-plane XRD pattern as well as with the earlier reports\cite{11,14}. The occupancies obtained from refinement for Dy and Sb are 1.0:0.9, which confirms the proper stoichiometry of grown crystal. All the refined parameters are tabulated in Table 1. Figure 2(d) represents the rock salt type crystal structue of DySb crystal generated from VESTA software, where Dy and Sb atoms are represented by Blue and brown spheres respectively. The Wychoff position for Dy and Sb are 4a(0,0,0) and 4b($\frac{1}{2}$,$\frac{1}{2}$,$\frac{1}{2}$) respectively.

\begin{figure}
	\centering
	\includegraphics[width=1\linewidth,height=0.45\linewidth]{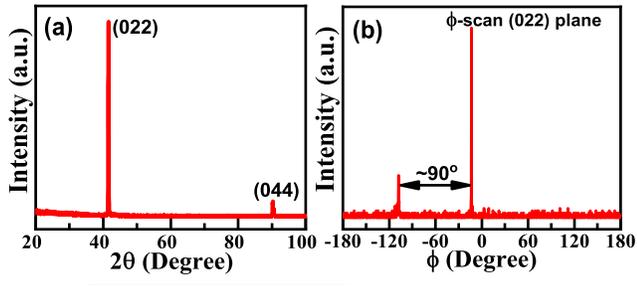}
	\caption{(a) $\theta$-2$\theta$, XRD scan of $\{$022$\}$ planes. (b) Azimuthal $\phi$-scan of (022) plane.}	
\end{figure}

\par
The out-of-plane XRD shows only $\{$002$\}$ set of planes i.e. $<$002$>$ oriented. To see the in-plane orientation of crystallographic planes, in-plane $\theta$-2$\theta$ XRD is performed for $\{$022$\}$ planes. Figure 3(a) shows the XRD pattern of DySb crystal along (022) plane. The XRD data exhibits only two peaks corresponding to (022) and (044) parallel planes. Hence, the grown crystal has a periodic arrangement of planes for out-of-plane [002] direction as well as in-plane [022] direction, confirming single crystalline nature of crystal. DySb crystallizes in NaCl-type structure therefore it has four-fold crystal symmetry, which means in azimuthal $\phi$-scan, one can observe a peak separation of $\sim$90\dg for four-fold crystal symmetry. Figure 3(b) shows the azimuthal $\phi$-scan for (022) crystal plane. The $\phi$-scan exhibits two peaks which are separated by $\sim$90\dg, suggesting the 4-fold (360\dg/90\dg=4) symmetry of the grown crystal. The observation of only sharp peaks in $\phi$-scan indicates that the grown crystal has perfect orientation of azimuthal domains.
\par

Figure 4(a) shows the rocking curve (RC) for (002) crystal plane. RC analysis is an important diagonastic tool to determine the crystalline perfection and disorder present in the crystal\cite{16,17}. The full width at half-maxima (FWHM), asymmetry and kink in RC play a crucial role for determination of crystalline disorders. The RC of (002) plane has a small kink at lower $\omega$ side of the main peak. To confirm this, RC (fig. 4(b)) is repeated for (004) plane where a similar type of kink is observed at lower $\omega$ side. This confirms that kink is inherently from the crystal and is well characterized. The deconvoluted RC of (002) peak is shown in fig. 4(c), which gives two peaks. The appearance of an additional peak at a separation of 0.03\dg from main crystal domain indicates a very low angle grain boundary in the internal structure. The high intense deconvoluted peak with FWHM$\sim$0.04\dg\, represents the main crystal domain while the low intense deconvoluted peak with FWHM$\sim$0.06\dg\, is appearing from very low angle grain boundaries having misorientation of 0.03\dg with the main crystal domain. The observed FWHM of main crystal domain in present work is smaller than that observed by Liu et al. in their DySb single crystal\cite{18}. Hence it is inferred that although the crystal has little defects, its small FWHM$\sim$0.04\dg of main crystal domain suggests a good crystalline nature of grown DySb single crystal\cite{16,19,20}.

\begin{figure}
	\centering
	\includegraphics[width=1\linewidth,height=0.55\linewidth]{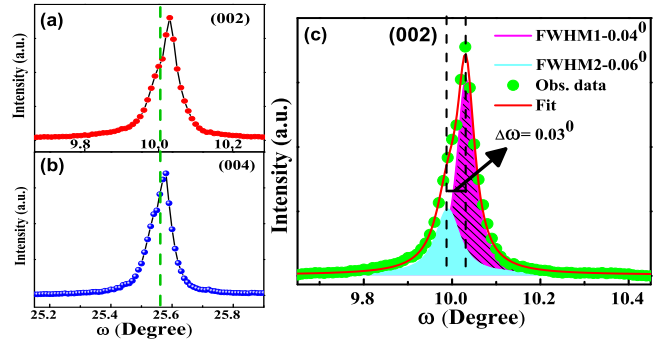}
	\caption{(a)-(b) Rocking curve (RC) of (002) and (004) plane of crystal, dotted line shows the presence of kink. (c) Deconvoluted RC of (002) diffracting plane.}	
\end{figure}

\subsection{Characterization of Phase transition by transport and magnetic measurements}

Figure 5(a) shows the temperature dependence of electrical resistance (R-T) of DySb single crystal in the absence of magnetic field. The zero-field resistance decreases gradually with decrease in temperature, demonstrating the metallic nature of DySb. A small kink is seen near 10\,K, whose enlarged view is shown in the upper inset of fig. 5(a). The enlarged view of R-T data shows a clear drop in resistance at  $\sim$9.7\,K showing PM to AFM transition. To find the appropriate transition point, the observed resistance data is differentiated with respect to temperature ($\frac{dR}{dT}$) which is shown in lower inset of fig. 5(a). The $\frac{dR}{dT}$ data shows a sharp peak at $\sim$9.7\,K that is well matched with the transition temperature reported in literature on DySb single crystals\cite{5,11,12,13,15}. We have further characterised our grown crystal by magnetisation measurements. Figure 5(b) illustrates the temperatuere dependence of magnetisation (M-T) for DySb crystal in an applied magnetic field of 0.05\,T. The M-T curve shows a sharp peak at $\sim$9.7\,K corresponding to the PM to AFM transition. This confirms that the synthesized single crystal grown by self-flux method have has a good crystalline quality.

\section{Conclusion}

\begin{figure}
	\centering
	\includegraphics[width=1\linewidth,height=0.45\linewidth]{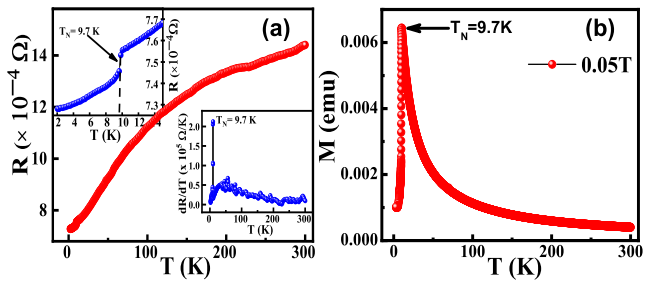}
	\caption{(a) Variation of electrical Resistance (R) as a function of temperature (T). Upper inset: Enlarged view of RT data below 15\,K showing transition point T$_N$. Lower inset: $\frac{dR}{dT}$ data as a function of temperature. (b) Temperature dependence of magnetisation (M-T) in an applied field of 0.05\,T.}	
\end{figure}

The shiny single crystal of DySb has been successfully grown by self-flux method.The Rietveld refinement of powder XRD pattern at room-temperature reveals the single-phase nature while HRXRD confirms its single crystalline nature. The crystalline system was identified to be cubic Fm$\bar{3}$m space group. Rocking curve analysis confirms the high quality of grown crystal. Phase transition temperature is confirmed by PM to AFM transition at T$_N$=9.7\,K in temperature dependent electrical transport as well as by magnetisation measurements, which further acertains the good quality of crystal.                                              

\begin{acknowledgements}
	
		We thank M. Gupta and L. Behera for XRD, V. R. Reddy and A. Gome for HRXRD and R. Vankatesh for FESEM merasurements. Dr. Devendra Kumar is acknowledged for fruitful discussions.
	
\end{acknowledgements}


\end{document}